\documentclass[%
reprint,
superscriptaddress,
 amsmath,amssymb,
 aps,
 physrev,
]{revtex4-1}

\usepackage{graphicx}
\usepackage{dcolumn}
\usepackage{bm}
\usepackage{physics}
\usepackage{siunitx}
\usepackage{mathtools}
\usepackage{extarrows} 

\usepackage{nth}

\usepackage{dsfont}

\usepackage{xr-hyper}
\usepackage{hyperref}
\hypersetup{
    colorlinks = true,
    allcolors = {blue},
    urlcolor = {black}
}

\usepackage{xfrac}

\usepackage{xcolor}

\usepackage{soul}
\usepackage{bigints}
\usepackage{subfigure}
\usepackage{chemformula}
\usepackage{lipsum} 
\usepackage[capitalize]{cleveref}

\def\permille{\ensuremath{{}^\text{o}\mkern-5mu/\mkern-3mu_\text{oo}}}

\NewDocumentCommand{\evalat}{sO{\big}mm}{%
  \IfBooleanTF{#1}
   {\mleft. #3 \mright|_{#4}}
   {#3#2|_{#4}}%
}

\newcommand{\appropto}{\mathrel{\vcenter{
  \offinterlineskip\halign{\hfil$##$\cr
    \propto\cr\noalign{\kern2pt}\sim\cr\noalign{\kern-2pt}}}}}

\begin{document}

\title{Quantum-enhanced absorption spectroscopy with bright squeezed frequency combs}

\author{Alexandre Belsley}
\email{alex.belsley@bristol.ac.uk}
\affiliation{%
Quantum Engineering Technology Labs, H. H. Wills Physics Laboratory and Department of Electrical \& Electronic Engineering, University of Bristol, Bristol BS8 1FD, United Kingdom
}%
\affiliation{%
Quantum Engineering Centre for Doctoral Training, H. H. Wills Physics Laboratory and Department of Electrical \& Electronic Engineering, University of Bristol, Bristol BS8 1FD, United Kingdom
}%

\date{\today}

\begin{abstract}
Absorption spectroscopy is a widely used technique that permits the detection and characterization of gas species at low concentrations. We propose a sensing strategy combining the advantages of frequency modulation spectroscopy with the reduced noise properties accessible by squeezing the probe state. A homodyne detection scheme allows the simultaneous measurement of the absorption at multiple frequencies and is robust against dispersion across the absorption profile. We predict a significant enhancement of the signal-to-noise ratio that scales exponentially with the squeezing factor. An order of magnitude improvement beyond the standard quantum limit is possible with state-of-the-art squeezing levels facilitating high precision gas sensing.
\end{abstract}
\maketitle

Spectroscopy is a precise and versatile tool to probe matter with key applications in process control~\cite{Lackner2007Apr}, chemical analysis~\cite{Bakker2002Jun}, and environmental monitoring~\cite{Laj2009Oct, Rieker2014Nov}. Accurate characterization of a gas phase absorption profile provides important physical information on gas composition, temperature, pressure, and velocity~\cite{Burgess1995Oct}.

Direct absorption spectroscopy can be performed using high-resolution, tunable laser diodes in the near and mid-infrared~\cite{Reid1978Jan, Hanson1978Aug, Werle2002Feb}. This technique, although common, is susceptible to laser intensity fluctuations, low-frequency noise sources, and spurious interference effects that limit the signal-to-noise (SNR) ratio and complicate the analysis of acquired spectra~\cite{Hodgkinson2012Nov}. 

An alternative to sweeping across the absorption spectrum is to frequency modulate a narrowband continuous-wave laser. In the limit of weak modulation, as first proposed by Bjorklund in 1979~\cite{Bjorklund1980Jan}, two weak sidebands can be created around the carrier frequency. The sideband frequency can be easily tuned by adjusting the modulation frequency, providing a convenient means to sweep probe light across an absorption feature.  Furthermore, orders of magnitude larger spectral resolution than is typical with a visible or near-infrared spectrometer can be obtained. 

Nonclassical states of light promise enhanced SNRs in a variety of sensing schemes~\cite{Dorfman2016Dec, Degen2017Jul, Lawrie2019Jun, Polino2020Jun}. Yurke and Whittaker~\cite{Yurke1987Apr} realized that by squeezing the two sidebands of a weakly frequency-modulated probe, the sensitivity of Bjorklund’s frequency modulation scheme could be improved beyond the standard quantum limit. An early experimental verification by Polzik \textit{et al.} yielded a 3.1 dB sensitivity improvement when probing the D2 line of atomic cesium~\cite{Polzik1992Sep}.

\begin{figure}[ht!]
    \centering
    \includegraphics[width=\linewidth]{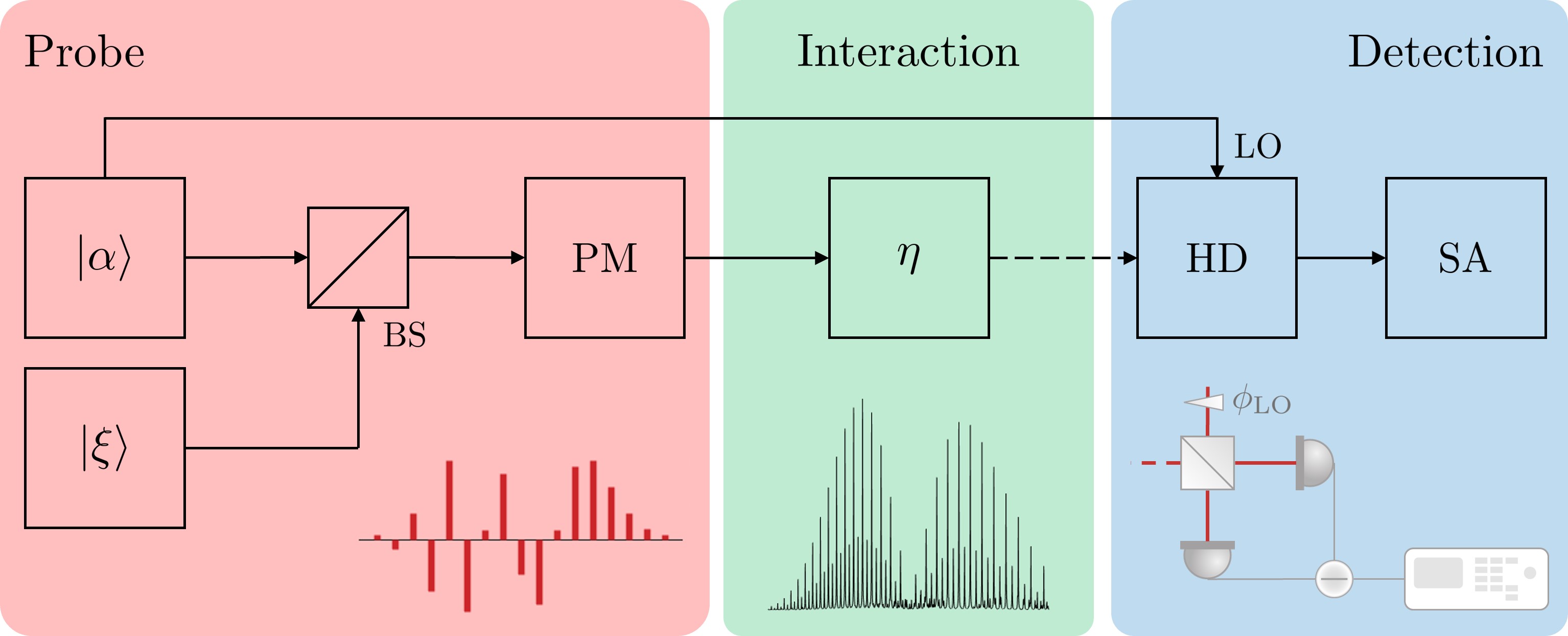}
    \caption{Schematic of the sensing strategy where a squeezed comb probes a gas with a frequency-dependent transmission $\eta$. The probe is generated by displacing a broadband squeezed vacuum state $\ket{\xi}$ with a coherent state $\ket{\alpha}$ using a highly reflective beamsplitter (BS) followed by a phase modulator (PM). The detection system consists of a balanced homodyne detector (HD) with a phase-tunable local oscillator (LO) followed by a spectrum analyzer (SA).}
    \label{fig:sketch}
\end{figure}

Here, we propose a sensing strategy that extends the advantages of using squeezed light in the weak modulation regime to a broader frequency comb generated by resorting to a higher modulation depth. This enables one to sample the wider absorption profiles of molecules at a dense, discrete set of frequencies without the need for scanning the incident laser or the modulation frequency. Instead of resorting to direct detection as in Yurke and Whittaker’s scheme, we propose the use of homodyne detection combined with an optical spectrum analyzer. This detection scheme amplifies the signal by the amplitude of the local oscillator (LO) field and the impact of the associated shot noise is avoided by subtracting the two photodiode currents~\cite{Yuen1983Mar}. Additionally, the transmission at the various sidebands can be measured simultaneously. In the limit of high transmission associated with low gas concentrations, we find that the SNR scales exponentially with the squeezing factor. For squeezing levels of 10 dB, an order of magnitude enhancement beyond the standard quantum limit is predicted. 

The sensing strategy we consider is schematically represented in~\cref{fig:sketch}. Broadband squeezed vacuum is displaced by a coherent state and then phase modulated to produce a squeezed frequency comb. This probe state interacts with a gas whose absorption properties we wish to characterize. The transmitted field is then detected using a homodyne detector followed by a spectrum analyzer.

To understand the scheme, it is helpful to initially focus our attention on the probe field at the carrier frequency $\omega_c$. This is a bright squeezed state, obtained by displacing the component of broadband squeezed vacuum $\ket{\xi}$ at $\omega_c$ with a coherent state $\ket{\alpha}$, i.e.,
\begin{equation}
    \alpha (\omega_c) + \hat{a}_s(\omega_c)\,.
\label{eq:mode_squeezer}    
\end{equation}
\noindent Here, $\alpha(\omega_c)$ is a classical coherent wave and $\hat{a}_s(\omega_c)$ is the annihilation operator arising from the squeezing operation. In the Caves formalism~\cite{Loudon1987Jun}, the squeezed vacuum operator $\hat{a}_s(\omega_c) = \hat{a}(\omega_c)\cosh(s) - e^{2 i \theta_s} \hat{a}^\dagger(\omega_c) \sinh(s)$ where $s$ is the squeezing factor, $\theta_s$ the squeezing angle and $\hat{a}^\dagger(\omega_c) = [\hat{a}(\omega_c)]^\dagger$ the creation operator. Squeezed vacuum can be generated using spontaneous parametric down-conversion~\cite{Wu1986Nov} or four-wave mixing~\cite{Slusher1985Nov}.

This field is in turn phase modulated with a modulation frequency $\Omega$ and depth $M$. In the following, we assume the phase modulator to be ideal, i.e., lossless and with negligible dispersion over the bandwidth of the target absorption line. The action of the phase modulator on the field at $\omega_c$ produces a series of frequency sidebands $\omega_c \pm n \Omega$ while maintaining the nonclassical properties associated with $\hat{a}_s$, which is transformed according to $\hat{a}_s(\omega_c) \rightarrow \sum_n J_n(M)\, \hat{a}_s (\omega_c + n\Omega)$ where $J_n(M)$ are the Bessel functions of order $n$~\cite{Capmany2011Jan, Horoshko2018Nov}. For simplicity, we have assumed that the modulator has been appropriately tuned to eliminate a possible $n$-dependent phase.

\begin{figure*}[t]
    \centering
    \includegraphics[width=\linewidth]{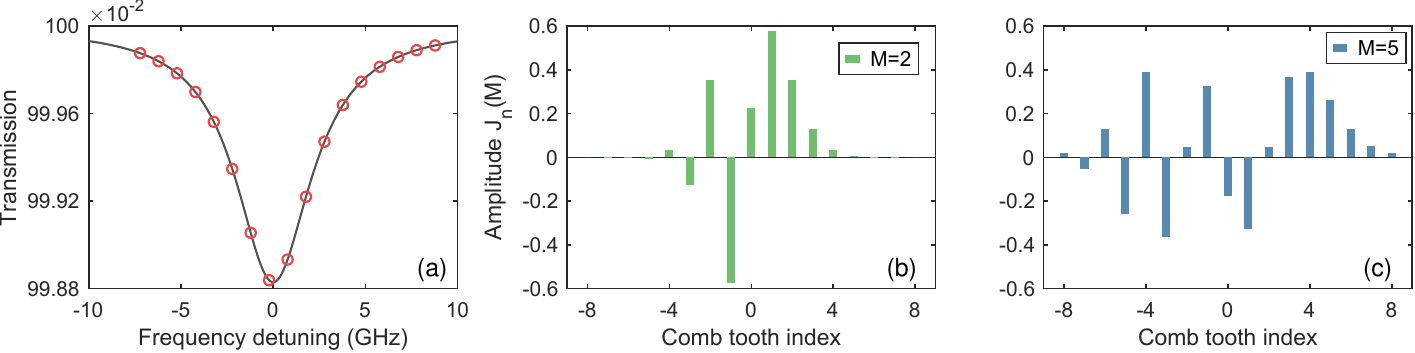}
    \caption{(a) Acetelyne's $\nu_1+\nu_3$ rotational-vibrational P9 transmission line (in black) with frequencies sampled by a 17-teeth comb with a modulation frequency $\Omega = \SI{1}{\GHz}$ (in red). Comb amplitudes for modulation depths (b) $M=2$ and (c) $M=5$.}
    \label{fig:transmission_comb_amplitudes}
\end{figure*}

The gas absorption can be modeled as a frequency-dependent beamsplitter with an amplitude transmission factor $\sqrt{\eta_{\pm n \Omega}}$, which simultaneously introduces a vacuum component with amplitude $\sqrt{1-\eta_{\pm n \Omega}}$. Because of dispersion about the absorption line, each frequency comb tooth also experiences a distinct optical phase shift $\phi_{\pm n \Omega}$.

The detection block consists of a balanced homodyne detector and a spectrum analyzer. The classical LO, $B_{\rm{LO}} = \abs{\beta} e^{i \phi_{\rm{LO}}}$, is derived from the initial coherent state field to ensure phase coherence. The LO phase $\phi_{\rm{LO}}$ is tunable, allowing us to measure either quadrature of the signal. Finally, the photodetectors are assumed to be identical and have efficiency $\eta_d$ as well as a flat-frequency response over the target gas absorption line profile.

The final step is to measure the spectral power $\mathcal{S}$ of the homodyne detector signal using a spectrum analyzer. The relevant detection signals are at one of the comb teeth frequencies of the modulated coherent state. For a frequency bin centered on a pair of comb teeth at $\omega_c \pm n\Omega$, the spectrum analyzer computes the spectral power 
\begin{equation}
\langle \hat{\mathcal{S}} (n \Omega) \rangle = \frac{1}{2} \sum_{r=\pm n} \langle \hat{i}(\omega_c +r \Omega) \hat{i} (\omega_c +r \Omega)^\dagger \rangle,
\label{eq:spectral_power}
\end{equation}
\noindent where $\hat{i}(\omega_c \pm n\Omega)$ is the subtraction photocurrent at the frequency sidebands $\omega_c \pm n \Omega$. The photocurrent has two components -- a classical contribution $\mathcal{I}(n\Omega) = 2 \alpha(n\Omega) \cos \phi_{\rm{LO}}$ and a quantum part proportional to the quadrature operator $\hat{x}_s(n\Omega)$, taking the form~\cite{Lvovsky2015Jan}
\begin{equation}
    \hat{i}(n\Omega) = \abs{\beta} \left[\mathcal{I}(n\Omega) + \sqrt{2} \hat{x}_s(n\Omega) \right].
\label{eq:photocurrent_freq}
\end{equation}
\noindent  For brevity, the frequency arguments have been referenced to the carrier frequency. In addition to the squeezed vacuum component at the carrier frequency $\omega_c$ in~\cref{eq:mode_squeezer}, the quadrature operator $\hat{x}_s$ also contains contributions of the broadband squeezed vacuum that were shifted by the phase modulator into the $\omega_c \pm n\Omega$ sidebands. The quadrature operator is thus given by
\begin{align}
    \hat{x}_s(n\Omega) = \frac{1}{\sqrt{2}}&\Big[\sqrt{\eta_{ n\Omega}}\,e^{i\phi_{n\Omega}} \sum_k J_{n-k}(M)\, \hat{a}_s\left((n-k)\Omega\right) \nonumber \\ &+ \sqrt{1-\eta_{n\Omega}}\, \hat{a}_\text{vac}(n\Omega)\Big] e^{-i\phi_{\text{LO}}}+\text{h.c.}\,,
\end{align}
\noindent where $\hat{a}_\text{vac}$ is the vacuum operator that arises from the interaction with the absorbing gas and $\text{h.c.}$ denotes the Hermitian conjugate. 
The effect of detector inefficiency is considered in Appendix~\ref{sec:Appendix_B_total}. 

We now assume the experimentally challenging case of weak absorption such that the difference in dispersion contributions $\Delta \phi = \phi_{+n \Omega} - \phi_{-n\Omega} \ll 1$. Under this assumption, the classical component of the photocurrent, $\mathcal{I} (\omega_c \pm n \Omega)$, leads to a normalized spectral density (see Appendix~\ref{sec:Appendix_A_classical})
\begin{align}
    \frac{\abs{I_{n\Omega}}^2}{\kappa^2}= &\cos^2(\Delta \phi_{\rm{LO}})\left[\sqrt{\eta_{+n\Omega}} + (-1)^n \sqrt{\eta_{-n\Omega}}\right]^2 \nonumber \\ &+ \sin^2(\Delta \phi_{\rm{LO}})\left[\sqrt{\eta_{+n\Omega}} - (-1)^n \sqrt{\eta_{-n\Omega}}\right]^2 \nonumber \\ &+2 (-1)^n\sin(2 \Delta \phi_{\rm{LO}}) \Delta\phi  \sqrt{\eta_{+n\Omega}\, \eta_{-n\Omega}} \,,
\label{eq:spectral_power_class}
\end{align}
\noindent where the normalization constant $\kappa\coloneqq \sqrt{\eta_d} \abs{\alpha\, \beta } J_n(M)$ and $\Delta\phi_{\rm{LO}} \coloneqq \phi_{n\Omega} - \phi_{\rm{LO}}$.

In general, this signal is proportional to a combination of the sum and difference between the transmission at complementary sidebands. The maximum and minimum spectral powers occur at LO phases $\Delta \phi_{\rm{LO}} = m \pi \lor (m+1) \pi/2, \,m \in \mathbb{Z}$. These extrema correspond to measuring the amplitude $X$ and phase $P$ quadratures. Note that under these two conditions, the third line of \cref{eq:spectral_power_class} that contains the dispersion contribution $\Delta \phi$ vanishes. In practice, the $X$ and $P$ quadratures can be identified by sweeping the LO phase through $2\pi$ radians. This readily allows one to separately obtain the transmission coefficients $\sqrt{\eta_{+n \Omega}}$ and $\sqrt{\eta_{-n \Omega}}$ at each sideband~\cite{commsJake}. As derived in Appendix~\ref{sec:Appendix_A_classical}, the amplitude transmission coefficient
\begin{equation}
\sqrt{\eta_{\pm n \Omega}} =  \frac{\abs{I_X \pm I_P} }{2 \kappa }\,,
\label{eq:transmission_eta}
\end{equation}
\noindent where the amplitude and phase quadrature contributions are respectively given by
\begin{align}
    I_X &= \abs{I_{n\Omega}} \Big|_{\Delta\phi_{\rm{LO}} = m \pi} \,, \label{eq:I_X_quadrature}\\
    I_P &= \abs{I_{n\Omega}} \Big|_{\Delta\phi_{\rm{LO}} = (m+1) \pi/2}\label{eq:I_P_quadrature} \,.
\end{align}

A straightforward calculation of the quantum contribution leads to the following expression for the total spectral power at the sidebands $n \Omega$ (see Appendix~\ref{sec:Appendix_B_total})
\begin{align}
\langle \hat{\mathcal{S}} (n \Omega) \rangle=& \abs{I_{n\Omega}}^2 +  \abs{\beta}^2\eta_d \left[1+J_n^2 (M)\right](\eta_{+n\Omega} + \eta_{-n\Omega}) e^{-2s} \nonumber \\ & +\abs{\beta}^2 \eta_d (2-\eta_{+n\Omega}-\eta_{-n\Omega}) +  (1-\eta_d)\,.
\label{eq:spectral_power_total}
\end{align}
\noindent For a bright input coherent state and a strong LO, i.e., $\abs{\alpha}^2, \abs{\beta}^2 \gg 1$, the classical term $\abs{I_{n\Omega}}^2$ whose explicit form is given in~\cref{eq:spectral_power_class} dominates.

After a lengthy calculation, we obtain the spectral power variance (see Appendix~\ref{sec:Appendix_C_variance})
\begin{align}
    &\Delta^2 \mathcal{S} (n\Omega) = 2\,\eta_d^2\, \abs{\beta}^4 \big\{[1+J_n^2(M)]^2 (\eta_{+n\Omega} +\eta_{-n\Omega})^2\, e^{-4s} \nonumber \\ &+2\left[1+J_n^2(M)\right](\eta_{+n\Omega}+\eta_{-n\Omega})(2-\eta_{+n\Omega}-\eta_{-n\Omega})e^{-2s} \nonumber \\ & +(2-\eta_{+n\Omega}-\eta_{-n\Omega})^2\big\} + 2  (1-\eta_d)^2\,.
\label{eq:spectral_power_var}
\end{align}
\noindent By virtue of using a homodyne detector with a strong LO, the noise is dominated by the losses associated with the finite transmission of the sample. For low gas absorption $\eta_{\pm n \Omega} \lessapprox 1$, the first term in \cref{eq:spectral_power_var} dominates. The quantum advantage scales exponentially with the squeezing factor, that is $\Delta \mathcal{S}_{\text{SQL}} / \Delta \mathcal{S} \gtrsim e^{2s}$ where the standard quantum limit $\Delta \mathcal{S}_{\text{SQL}}$ is obtained by setting $s=0$. For a squeezing level of 13 dB ($s \approx 1.5$) measured at \SI{1550}{\nm}~\cite{Schonbeck2018Jan}, this translates to a precision enhancement by a factor of 20.

As an example, we consider probing acetelyne's $\ch{C2H2}$ $\nu_1 + \nu_3$ rotational–vibrational band around \SI{1530}{\nm}. Acetylene is highly combustible and has industrial applications in welding and cutting. In \cref{fig:transmission_comb_amplitudes} (a), we plot the transmission profile of the $P9$ line for a path length $L = \SI{1}{\cm}$ and a partial pressure of 1$\permille$ at a total pressure of 1 atm using data from the HITRAN database~\cite{Rothman2013Nov}. The frequencies sampled by the central 17 teeth of a comb with a modulation frequency $\Omega = \SI{1}{\GHz}$ are also shown. 

The Bessel function dependence on the modulation depth allows one to make the signal at certain sidebands more prominent by varying $M$. This is illustrated in \cref{fig:transmission_comb_amplitudes} (b) and (c)  for two different modulation depths $M=2$ and $M=5$ where we see that, in general, the strongest teeth extend roughly from $-M$ to $M$. This is a reflection of the Bessel function's dependence on $M$ known as Carson’s rule~\cite{Carson1922Feb}. Note that at higher modulation depths (e.g. $M=5$), some of the lower-order teeth have a reduced amplitude. In practice, by repeating the experiment at different values of $M$, one can obtain a higher SNR across the transmission profile. Additionally, the comb teeth density can be tuned by adjusting the modulation frequency $\Omega$.
\begin{figure*}[t!]
    \centering
    \includegraphics[width=\linewidth]{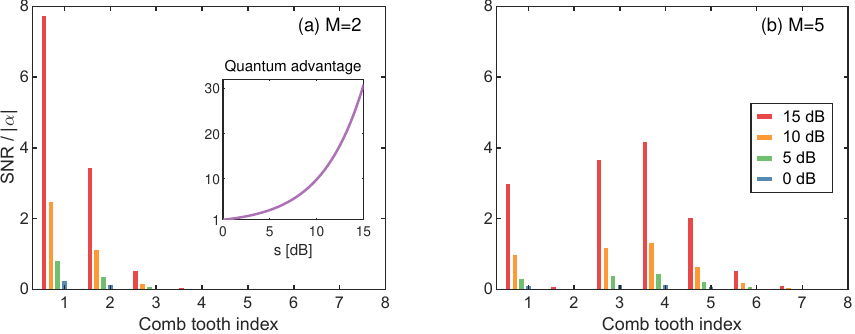}
    \caption{Signal-to-noise ratio $\langle \hat{\mathcal{S}} (n \Omega) \rangle/ \Delta \mathcal{S}(n \Omega)$ normalized by the input-coherent state amplitude $\abs{\alpha}$ for different squeezing levels when probing acetylene's P9 line at 1$\permille$ partial pressure and a $\SI{1}{\cm}$ path length. A frequency comb with 17 teeth, modulation frequency $\Omega = \SI{1}{\GHz}$ and depth (a) $M=2$ or (b) $M=5$ was used. Inset: quantum advantage in the SNR as a function of the squeezing level $s$. We have assumed ideal detection and a bright local oscillator as described in the main text.}
    \label{fig:SNR}
\end{figure*}

In \cref{fig:SNR}, we plot the predicted normalized SNR, $\langle \hat{\mathcal{S}} (n \Omega) \rangle/ \Delta \mathcal{S}(n \Omega)$, for modulation depths $M=\{2,5\}$ under ideal conditions where $\eta_d=1$ and there are no additional technical noise sources. To normalize, we divided the SNR by the coherent state amplitude under the assumptions that $\abs{\alpha}^2 \gg \sinh^2(s)$ and the LO amplitude $\beta \gg 1$. In practice, the SNR scales directly with $\abs{\alpha}$ such that even a modestly bright coherent state will give rise to a large SNR. For the largest squeezing level considered ($s=\SI{15}{dB}$), the quantum advantage in the SNR reaches a factor of 31. At the higher modulation depth $M=5$, the squeezing advantage in the SNR at the $n=\pm 2$ teeth is reduced due to the small amplitude of the Bessel function for these teeth. Nevertheless, by using a lower modulation depth $M=2$, these central teeth can be measured with a large SNR.

In summary, we have proposed a quantum-enhanced absorption spectroscopy method where a given gas is probed by a bright squeezed frequency comb. We predict an enhancement of the signal-to-noise ratio by an order of magnitude at squeezing levels of 10 dB. This would allow one to detect the presence of environmental gases at concentrations that are an order of magnitude lower than with a classical input probe.  Remarkably, for weak absorption when the quantum advantage is greatest, the proposed measurement scheme is robust against dispersion effects across the absorption profile.

In essence, this quantum enhancement arises from the reduced amplitude uncertainty in the input broadband squeezed vacuum. While the phase modulator redistributes this field into the various comb teeth, its nonclassical properties are retained provided the squeezer is sufficiently broadband and the phase modulator has negligible dispersion and loss over the target absorption line.

The spectral density of the transmitted signal is measured using a homodyne detector and an optical spectrum analyzer. The transmission of the gas at each frequency sampled by a given comb tooth can be independently determined by simply tuning the local oscillator phase. We thus avoid the need for multiple sequential measurements as in cavity ring-down
spectroscopy~\cite{Berden2000Oct,Thorpe2006Mar} or the long acquisition times present in conventional Fourier-transform infrared spectrometers~\cite{Griffiths2007Apr}. We also do not require high-resolution dispersive spectrometers~\cite{Diddams2007Feb,Gohle2007Dec,Karim2020Aug} or Fourier transform spectrometer techniques with long delay lengths~\cite{Maslowski2016Feb} commonly used in direct frequency comb spectroscopy~\cite{Stowe2008Jan}. Moreover, in our method, the offset of the carrier frequency from absorption line center can also be readily estimated from the asymmetry in the measured absorption at the different comb teeth pairs.

Another key advantage of our scheme is the inherent mutual coherence between the frequency comb teeth and the local oscillator. This is in contrast to dual-comb spectroscopy methods, where it is nontrivial to ensure mutual coherence between the two combs~\cite{Coddington2016Apr, Picque2019Mar}. Furthermore, the ability to vary the modulation depth allows sampling of both the wings and the central portion of the absorption spectrum with roughly the same, large signal-to-noise ratio. Practically, this enables one to accurately determine the line shape and strength of an isolated transition and to infer properties such as the concentration, temperature, or pressure of a given gas species.

The strategy we propose also has significant advantages over quantum-enhanced absorption spectroscopy strategies using multiphoton entangled probes~\cite{Dinani2016Jun}. These probe states are nontrivial to generate and their performance quickly degrades in the presence of external system losses. Alternative sensing strategies using Fock~\cite{Adesso2009Apr, Whittaker2017Feb} or two-mode squeezed vacuum~\cite{Shi2020Oct} state probes are more robust to external losses. However, compared to our proposal, these probe states cannot be readily generated with macroscopic photon numbers and require considerably more sophisticated detection schemes, which involve optical parametric amplification before photodetection or photon counting detectors.

Finally, the sensing setup we propose has no moving parts and could be realized in fiber using standard components. Recent advances in integrated frequency comb sources~\cite{Gaeta2019Mar, Xiang2021Jul, Yang2021Aug, Hu2022Aug}, low-loss waveguide-based sensing structures~\cite{Puckett2021Feb, Belsley2022Jun} and on-chip homodyne detectors~\cite{Tasker2021Jan} suggest the scheme proposed could also be fully integrated, leading to a compact and robust spectroscopic gas sensor.

\begin{acknowledgements}
A.B. acknowledges support from the ERC starting grant ERC-2018-STG 803665 and the EPSRC grant EP/S023607/1. All the data needed to evaluate the conclusions of the paper are present in the main text and in the Appendix.
\end{acknowledgements}

\bibliography{mainbib.bib}

\clearpage
\onecolumngrid
\begin{center}
\textbf{\large APPENDIX}
\end{center}
\setcounter{equation}{0}
\setcounter{figure}{0}
\setcounter{table}{0}
\renewcommand{\theequation}{A\arabic{equation}}
\renewcommand{\thefigure}{A\arabic{figure}}
\setcounter{page}{1}

Here, we supplement the main text by including further details on the derivation of \cref{eq:spectral_power_class,eq:I_X_quadrature,eq:I_P_quadrature,eq:spectral_power_total,eq:spectral_power_var}.

\subsection{Classical spectral power}
\label{sec:Appendix_A_classical}

We start with the classical coherent field $\alpha(t) = \abs{\alpha} e^{i \omega_c t}$ at the carrier frequency $\omega_c$. After the phase modulator, we obtain $\alpha_{\text{PM}}(t) = \abs{\alpha} \sum_{n} J_n(M) e^{-i(\omega_c + n \Omega) t}$~\cite{Bjorklund1980Jan}. This field can be written as a central frequency component at $\omega_c$ together with a series of upper and lower sidebands at frequencies $\omega_c \pm n\Omega$, that is
\begin{align}
    \alpha_{\text{PM}}(t) = \abs{\alpha} \Big[J_0 (M) e^{-i \omega_c t} + \sum_{n=1}^{\infty} J_n(M)  e^{-i (\omega_c+n \Omega) t} +  \sum_{n=1}^{\infty} (-1)^n J_n(M) e^{-i (\omega_c-n \Omega) t}\Big].
\label{eq:mode_FM}
\end{align}
\noindent Following the interaction with the gas, the field components at a pair of sidebands $\omega_c \pm n \Omega$ take the form
\begin{align}
        \alpha_{\text{G}}(t) = \abs{\alpha} \Big[\sqrt{\eta_{+n\Omega}}\, e^{-i \phi_{+n\Omega}} J_n(M)  e^{-i (\omega_c+n \Omega) t}  +  \sqrt{\eta_{-n\Omega}}\, e^{-i \phi_{-n\Omega}} (-1)^n J_n(M) e^{-i (\omega_c-n \Omega) t} \Big].
\label{eq:coherent_post_gas}
\end{align}
\noindent For a classical signal field $\alpha(t)$, the subtraction photocurrent at the homodyne detector output is given by~\cite{Lvovsky2015Jan}
\begin{align}
    i(t) = \int d\tau\, k(t-\tau) \left[ \alpha(\tau) B_{\rm{LO}}^*(\tau)+\text{c.c.}\right] = \int d\tau\, k(t-\tau) \sqrt{\eta_d} \abs{\beta}  \left[\alpha(\tau) e^{-i\phi_{\text{LO}}}+\text{c.c.}\right],
\label{eq:photocurrent_t}
\end{align}
\noindent where $k(t)$ is the detector response function, assumed to be lossless and instantaneous over the gas absorption line, and c.c. denotes the complex conjugate. Substituting~\cref{eq:coherent_post_gas} into~\cref{{eq:photocurrent_t}} leads to the subtraction photocurrent
\begin{align}
\mathcal{I} (t) = &\,2 \sqrt{\eta_d} \abs{\alpha \beta} J_n(M) \big\{ \cos(n \Omega t)    \left[\sqrt{\eta_{+n \Omega}} \cos(\phi_{+n\Omega}-\phi_{\text{LO}}) + (-1)^n \sqrt{\eta_{-n \Omega}} \cos(\phi_{-n\Omega}-\phi_{\text{LO}})\right] \nonumber \\ &+  \sin(n \Omega t) \left[\sqrt{\eta_{+n \Omega}} \sin(\phi_{+n\Omega}-\phi_{\text{LO}}) - (-1)^n \sqrt{\eta_{-n \Omega}} \sin(\phi_{-n\Omega}-\phi_{\text{LO}})\right]\big\}.
\end{align}
\noindent Defining $\Delta\phi_{\text{LO}} \coloneqq \phi_{+n\Omega}-\phi_{\text{LO}}$ and $\kappa \coloneqq \sqrt{\eta_d} \abs{\alpha \beta} J_n(M)$, this equation can be written as
\begin{align}
\frac{\mathcal{I}(t)}{2 \kappa} = & \cos(n \Omega t) \left\{ \cos(\Delta\phi_{\text{LO}})    \left[\sqrt{\eta_{+n \Omega}}  + (-1)^n \sqrt{\eta_{-n \Omega}}\right] + \sin(\Delta\phi_{\text{LO}}) \Delta\phi (-1)^n \sqrt{\eta_{-n \Omega}}   \right\} \nonumber \\ & +\sin(n \Omega t) \left\{ \sin(\Delta\phi_{\text{LO}})    \left[\sqrt{\eta_{+n \Omega}}  - (-1)^n \sqrt{\eta_{-n \Omega}}\right] + \cos(\Delta\phi_{\text{LO}}) \Delta\phi (-1)^n \sqrt{\eta_{-n \Omega}} \right\},
\end{align}
\noindent where we have assumed weak absorption such that the dispersion between each pair of comb teeth $\Delta \phi = \phi_{+n \Omega} - \phi_{-n \Omega} \ll 1$. 

Taking the Fourier transform and computing the spectral power using \cref{eq:spectral_power} readily yields \cref{eq:spectral_power_class} of the main text, which is reproduced below for convenience
\begin{align}
    \frac{\abs{I_{n\Omega}}^2}{\kappa^2}= & \cos^2(\Delta \phi_{\rm{LO}})\left[\sqrt{\eta_{+n\Omega}} + (-1)^n \sqrt{\eta_{-n\Omega}}\right]^2 + \sin^2(\Delta \phi_{\rm{LO}})\left[\sqrt{\eta_{+n\Omega}} - (-1)^n \sqrt{\eta_{-n\Omega}}\right]^2 \nonumber \\ &+ 2 (-1)^n\sin(2 \Delta \phi_{\rm{LO}}) \Delta\phi  \sqrt{\eta_{+n\Omega}\, \eta_{-n\Omega}} \,.
\label{eq:dispersion_S1}
\end{align}
In the zero dispersion limit where $\Delta\phi =0$, the maximum and minimum spectral power are easily verified to occur at $\Delta \phi_{\rm{LO}} = m \pi \lor (m+1)\pi/2, \,m \in \mathbb{Z}$. This provides a simple experimental means to identify the two quadrature contributions accessible by sweeping the LO phase. Note that for these two LO phases, the third term in \cref{eq:dispersion_S1} is identically zero allowing us to readily identify the amplitude and phase quadratures contributions given in \cref{eq:I_X_quadrature,eq:I_P_quadrature} of the main text, namely
\begin{align}
    I_X &= \abs{I_{n\Omega}} \Big|_{\Delta\phi_{\rm{LO}} = m \pi} = \abs{\kappa \cos(\Delta \phi_{\rm{LO}})\left[\sqrt{\eta_{+n\Omega}} + (-1)^n \sqrt{\eta_{-n\Omega}}\right]}\,, \\
    I_P &= \abs{I_{n\Omega}} \Big|_{\Delta\phi_{\rm{LO}} = (m+1) \pi/2} = \abs{\kappa \sin(\Delta \phi_{\rm{LO}})\left[\sqrt{\eta_{+n\Omega}} - (-1)^n \sqrt{\eta_{-n\Omega}}\right]}\,.
\end{align}
It is possible to infer whether the central tooth is tuned to line center of a symmetric absorption profile by observing whether the extrema of the signal switch between the $X$ and $P$ quadratures for adjacent comb teeth.

The influence of the dispersion on the maximum and minimum spectral power signals can be readily obtained by calculating the null points of the derivative of \cref{eq:dispersion_S1} with respect to $\Delta\phi_{\rm{LO}}$, yielding
\begin{equation}
    \Delta\phi_{{\rm{LO}}_{\text{max/min}}} = \frac{1}{2} \atan( \frac{2 \Delta \phi}{1+\sqrt{\eta_{+n\Omega}/\eta_{-n\Omega}}})\,.
\end{equation}
\noindent Finally, by Taylor expanding around $\Delta\phi=0$, we obtain $\Delta\phi_{{\rm{LO}}_{\text{max/min}}} = m \frac{\pi}{2} + \frac{\Delta \phi}{1+\sqrt{\eta_{+n\Omega}/\eta_{-n\Omega}}}$. For low absorption, the phase difference accumulated between teeth pairs $\Delta \phi \ll 1$ such that $\Delta\phi_{{\rm{LO}}_{\text{max/min}}} \approx m \pi/2$ to a good approximation, i.e. the zero dispersion limit. Consequently, provided that the dispersion difference between symmetric teeth pairs $\Delta\phi \leq 0.01$ , the change in the local oscillator phase $\Delta\phi_{{\rm{LO}}_{\text{max/min}}}$ that gives rise to a maximum and minimum signal compared to the zero dispersion limit is less than $1/100\textsuperscript{th}$ of a radian.

\subsection{Total spectral power}
\label{sec:Appendix_B_total}

Accounting for detector inefficiency, the quadrature operator $\hat{x}_s$ in \cref{eq:photocurrent_freq} of the main text is given by
\begin{align}
    \sqrt{2}\, \hat{x}_s(n\Omega) = &\sqrt{\eta_d}\Big[\sqrt{\eta_{ n\Omega}}\,e^{i\phi_{n\Omega}} \sum_k  J_{n-k}(M)\, \hat{a}_s\left((n-k)\Omega\right) + \sqrt{1-\eta_{n\Omega}}\, \hat{a}_\text{vac}(n\Omega)\Big]e^{-i\phi_{\text{LO}}} \nonumber \\&+\sqrt{1-\eta_d}\, \hat{a}_{\text{vac}^\prime}(n\Omega) +\text{h.c.}\,.
\end{align}
\noindent Here, we have used the standard quantum optics model for detector inefficiency as a beamsplitter with a transmission coefficient $\sqrt{\eta_d}$, which also introduces a vacuum component $\hat{a}_{\text{vac}^\prime}$ with amplitude $\sqrt{1-\eta_d}$.

The subtraction photocurrent at frequency sidebands $\omega_c \pm n \Omega$ then takes the following form
\begin{align}
    \hat{i}(n\Omega) =\,&\mathcal{I} (n\Omega) +
    \abs{\beta} \sqrt{\eta_d} \Big[\sqrt{\eta_{ n\Omega}}\,e^{i\phi_{n\Omega}} \sum_k  J_{n-k}(M)\, \hat{a}_s\left((n-k)\Omega\right) + \sqrt{1-\eta_{n\Omega}}\, \hat{a}_\text{vac}(n\Omega)\Big]e^{-i\phi_{\text{LO}}} \nonumber \\&+\sqrt{1-\eta_d}\, \hat{a}_{\text{vac}^\prime}(n\Omega) +\text{h.c.}\,.
\end{align}
\noindent By assuming broadband squeezing where each relevant frequency component is quadrature squeezed by the same amount, it follows that $\langle \hat{x}_s(n\Omega)^2 \rangle = e^{-2s}$. Substituting $\hat{i}(n\Omega)$ into \cref{eq:spectral_power} results in the total spectral power given in \cref{eq:spectral_power_total} of the main text, i.e.,
\begin{align}
\langle \hat{\mathcal{S}} (n \Omega) \rangle= \abs{I_{n\Omega}}^2 +  \abs{\beta}^2 \eta_d  \left[1+J_n^2 (M)\right](\eta_{+n\Omega} + \eta_{-n\Omega}) e^{-2s} +\abs{\beta}^2 \eta_d (1-\eta_{+n\Omega}+1-\eta_{-n\Omega}) +  (1-\eta_d)\,.
\end{align}

\subsection{Spectral power variance}
\label{sec:Appendix_C_variance}

The variance in spectral power is $\Delta^2 \mathcal{S} (n\Omega) = \langle \mathcal{S}(n \Omega)^2\rangle - \langle \mathcal{S}(n\Omega)\rangle^2$ and a straightforward calculation of $\langle \mathcal{S}(n \Omega)^2\rangle$ yields
\begin{align}
    \langle \mathcal{S}(n \Omega)^2\rangle =  &\,3\,\eta_d^2\, \abs{\beta}^4 \big\{[1+J_n^2(M)]^2 (\eta_{+n\Omega} +\eta_{-n\Omega})^2\, e^{-4s} +2\left[1+J_n^2(M)\right](\eta_{+n\Omega}+\eta_{-n\Omega})(1-\eta_{+n\Omega}+1-\eta_{-n\Omega})e^{-2s} \nonumber \\ & +(1-\eta_{+n\Omega}+1-\eta_{-n\Omega})^2\big\} + 3(1-\eta_d)^2 + \abs{I_{n\Omega}}^4\,.
\end{align}
\noindent Using the expression for $\langle \mathcal{S}(n\Omega)\rangle$ given in \cref{eq:spectral_power_total} of the main text, the variance $\Delta^2 \mathcal{S}$ at sidebands with frequencies $\omega_c \pm n\Omega$ is found to equal
\begin{align}
    \Delta^2 \mathcal{S} (n\Omega) = &\,2\,\eta_d^2\, \abs{\beta}^4 \big\{[1+J_n^2(M)]^2 (\eta_{+n\Omega} +\eta_{-n\Omega})^2\, e^{-4s} +2\left[1+J_n^2(M)\right](\eta_{+n\Omega}+\eta_{-n\Omega})(1-\eta_{+n\Omega}+1-\eta_{-n\Omega})e^{-2s} \nonumber \\ & +(1-\eta_{+n\Omega}+1-\eta_{-n\Omega})^2\big\} + 2  (1-\eta_d)^2\,,
\end{align}
\noindent which corresponds to \cref{eq:spectral_power_var} of the main text.

\end{document}